\documentclass{article}
\usepackage{spconf,amsmath,graphicx}

\usepackage{enumitem}
\setlist{nosep, leftmargin=14pt}
\usepackage{graphics}

\usepackage{mwe} 

\def\A{{\cal A}}

\title{Unsupervised Anomaly Detection in Digital Pathology using GANs}
\name{Milda Pocevičiūtė$^{\star \dagger}$, Gabriel Eilertsen$^{\star \dagger}$, Claes Lundstr\"om$^{\star \dagger \mathsection}$}
\address{$^{\star}$Department of Science and Technology, Linköping University, Sweden\\
$^{\dagger}$Center for Medical Image Science and Visualization, Link\"oping University\\
$^{\mathsection}$Sectra AB, Linköping, Sweden
}
\begin{document}
%
\begin{minipage}[t]{0.9\textwidth}
\Huge IEEE Copyright Notice 
\newline

\small © 20XX IEEE.  Personal use of this material is permitted.  Permission from IEEE must be obtained for all other uses, in any current or future media, including reprinting/republishing this material for advertising or promotional purposes, creating new collective works, for resale or redistribution to servers or lists, or reuse of any copyrighted component of this work in other works.

\vspace*{5\baselineskip}
\large Accepted to be Published in: Proceedings of the 2021 IEEE International Symposium on Biomedical Imaging (ISBI), April 13-16, 2021.
\end{minipage}

\maketitle
\begin{abstract}

Machine learning (ML) algorithms are optimized
for the distribution represented by the training data. For outlier data, they often deliver predictions with equal confidence, even though these should not be trusted. In order to deploy ML-based digital pathology solutions in clinical practice, effective methods for detecting anomalous data are crucial to avoid incorrect decisions in the outlier scenario. We propose a new unsupervised learning approach for anomaly detection in histopathology data based on generative adversarial networks (GANs). 
Compared to the existing GAN-based methods that have been used in medical imaging, the proposed approach improves significantly on performance for pathology data.
Our results indicate that histopathology imagery is substantially more complex than the data targeted by the previous methods. This complexity requires not only a more advanced GAN architecture but also an appropriate anomaly metric to capture the quality of the reconstructed images.
\end{abstract}
\section{Introduction} \label{intro}

Machine learning, and in particular deep learning, is showing great potential 
in pathology applications, with state-of-the-art methods achieving pathologist-level performance~\cite{Bulten2020}. To guarantee reliable use of these approaches in practice, we have to detect the situations when the algorithm encounters outlier data and cannot be trusted. Hence, anomaly detection is an important topic in digital pathology, making it possible to flag potential mispredictions to the pathologist. 

A fundamental challenge in anomaly detection is that the anomaly is unknown or unavailable at training time, requiring an unsupervised training strategy. For this purpose, frameworks based on generative adversarial networks (GANs)~\cite{goodfellow2014} have shown great promise \cite{originalAnoGAN, fAnoGAN, am2019unsupervised}. As GANs learn to capture data distribution characteristics, they can be used to estimate if a sample fits the distribution or should be considered an outlier/anomaly. With the fast development in GAN architectures and training strategies, e.g. including Wasserstein GAN (WGAN)~\cite{gulrajani2017improved}, Progressive Growing GAN (PG--GAN)~\cite{karras2017progressive} and StyleGAN2~\cite{stylegan2}, the potential for GAN-based anomaly detection is constantly growing. For an in-depth introduction to the field of anomaly detection, we refer to a recent survey by Bulusu et al \cite{OOD_survey}.

In this work, we introduce s2-AnoGAN, a GAN-based unsupervised anomaly detection method. 
Our approach stems from two important observations for the histopathology case. First, a powerful GAN architecture is needed to capture the high data complexity. Second, anomaly scores relying on pixel-wise comparisons are not suitable for this application domain. Instead, we propose a new anomaly score that uses edge information. We show that in combination these two components allow for substantial improvements over existing methods when evaluated on pathology data.

\section{Methods and Materials} \label{sec:meth}
\begin{figure}[t]
    \begin{minipage}[b]{1.0\linewidth}
      \def\cpar{\hss\egroup\line\bgroup\hss}
      \centering
      \centerline{\includegraphics[width=8.0cm]{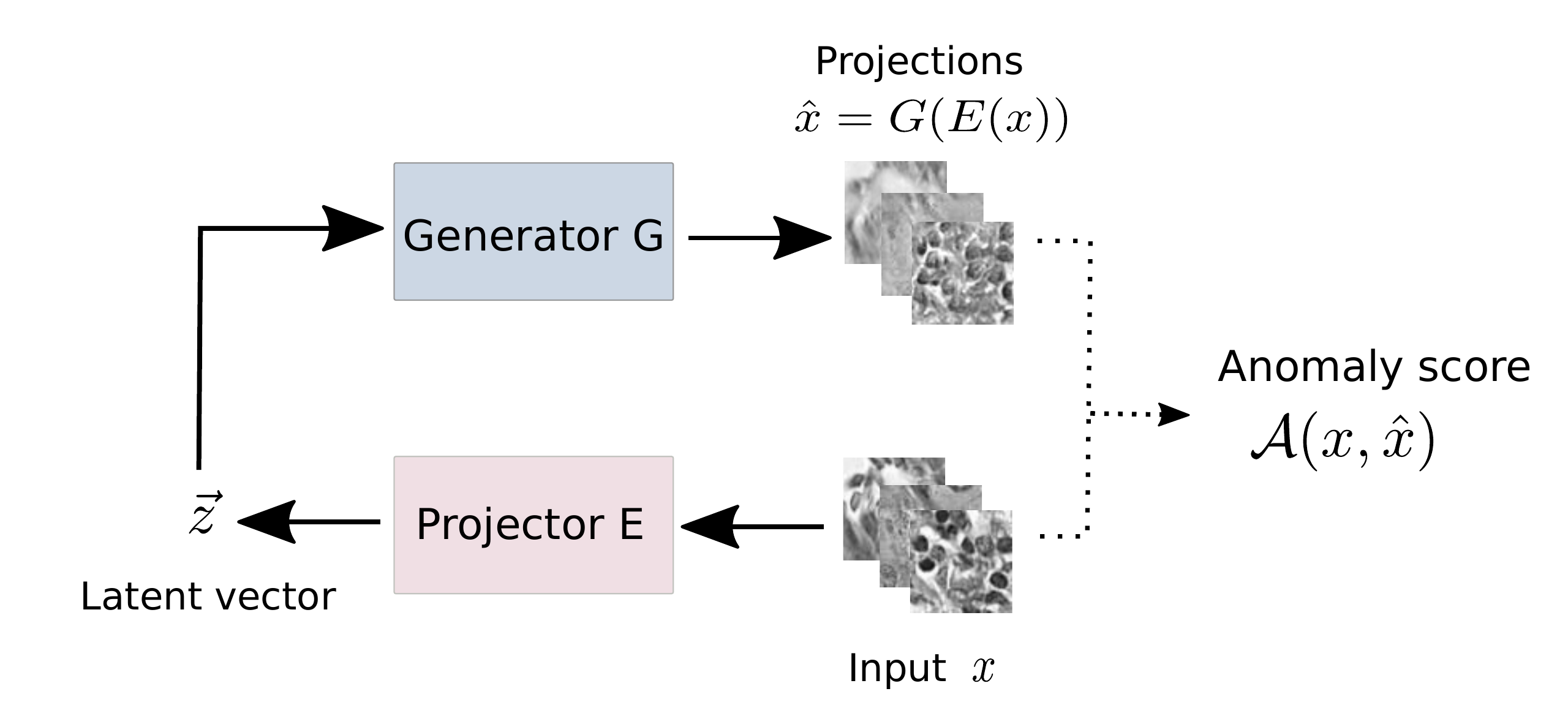}}
    \end{minipage}
    \centering
    \caption{Scheme of GAN-based unsupervised anomaly detection. The original input image $x$ is fed through projector E which outputs a latent vector $\vec{z}$. This latent vector is used to generate a reconstruction image via generator G. An anomaly score is a measure of some difference between the input image $x$ and its reconstruction $\hat{x}$.}
    \label{fig_general}
\end{figure} 

We evaluate the performance on pathology data of our proposed s2-AnoGAN method, and compare it to two previous methods: f-AnoGAN~\cite{fAnoGAN} and a framework introduced by Berg et al.~\cite{am2019unsupervised} which we denote pg-AnoGAN.

Fig.~\ref{fig_general} illustrates the general idea behind unsupervised anomaly detection based on GANs. Given a trained generator $G$, and a projector $E$ (estimated inverse of $G$), the anomaly score is a distance function $\A(x, \hat{x})$ that measures the difference between input image $x$ and its reconstructed projection $\hat{x} = G(E(x))$. The anomaly detection frameworks differ in a choice of GAN, projection method, and anomaly measure $\A$.

Our proposed s2-AnoGAN anomaly detector is based on the StyleGAN2 algorithm (configuration F) and its projector \cite{stylegan2}. The projector is running a gradient-based optimisation to minimise the LPIPS distance \cite{LPIPS} between the input image and its reconstruction. 
The f-AnoGAN framework is based on Wasserstein GAN with gradient penalty \cite{gulrajani2017improved} and the projector is trained separately using \textit{image-to-image} mapping approach. 
The pg-AnoGAN framework uses PG-GAN \cite{karras2017progressive} and designs the encoder inspired by \cite{ClusterGAN}. For more details, please refer to the aforementioned papers \cite{fAnoGAN}, \cite{am2019unsupervised} and \cite{stylegan2}. The training of each algorithm was stopped once the perceived quality of the generated images was not changing for several epochs, resulting in 50 training epochs for f-AnoGAN, and 80 epochs for the pg-AnoGAN and s2-AnoGAN frameworks. Following the original implementation, f-AnoGAN encoder was trained separately for 50,000 iterations. 

Another differing aspect of the three frameworks is the anomaly score used. Our proposed framework includes a new anomaly score defined by $\A_{Canny} = N_{x_{Canny}} - N_{\hat{x}_{Canny} }$, where $x_{Canny}$ and $\hat{x}_{Canny}$ are the output images by the Canny edge detector \cite{canny_edges} applied to the original images and their reconstructions, respectively. $N_{y}$ is the count of non-zero pixels in a binary image $y$. In our experiments, the parameters were set to kernel 5, sigma 3 as well as 100 and 200 for the lower and the upper thresholds, respectively. The intuition behind the $\A_{Canny}$ metric is that even if the exact position of the features in the reconstructed images may vary, the overall edge characteristics should be similar in the original patch and its reconstruction given that the projection was successful.

The anomaly score of f-AnoGAN is given by $\A_{f\text{-}AnoGAN} = \A_{MSE} + \kappa \cdot \A_{D}$, where $\A_{MSE}$ is a mean square error (MSE) between the original image and its reconstruction, $\A_{D}$ is the MSE between discriminator features of original image and its reconstruction, and $\kappa$ is a weight factor set to 1.
The anomaly score for pg-AnoGAN is defined as
$\A_{pg\text{-}AnoGAN} = \alpha \cdot \A_{res} + ( 1- \alpha) \cdot \A_0(z)$, where $\A_{res}$ is the $l_2$-norm between original image and its reconstruction after min-max normalisation, $ \A_0(z)$ is the distance to the origin of the latent vector, and $\alpha$ is a weight factor set to 0.05.

There are many possible types of anomalies in pathology applications. For our experiments, we adopted a setup where images of healthy tissue constituted normal data, whereas tumour images were considered as anomalies. While we do not propose this setup to represent a feasible anomaly detection scenario in the clinic as such, it was chosen as a suitable proxy task to evaluate anomaly detection methods. The difference between healthy and tumorous tissue stems from complex histology characteristics, and this is a main challenge for any anomaly detection scenario in our domain.

The data used for the experiments consists of two types of patches, healthy and tumourous tissue, extracted from the CAMELYON17 data set \cite{camelyon17}, which contains whole-slide images (WSI) of hematoxylin and eosin (H\&E) stained lymph node sections. Patches are 64$\times$64 pixels, matching the image size used in the f-AnoGAN paper \cite{fAnoGAN}. We worked with grayscale images, in order to make fair comparisons to previous results, and to reduce the effect of colour variation in the images due to being sourced from different hospitals and scanners. The training set consisted of 87 521 healthy patches, while the held-out test data contained 9 729 healthy and 9 729 tumour patches. For the implementation of f-AnoGAN, we also extracted 12 503 healthy patches for training the encoder.
\section{Results and Discussion} \label{sec:results}

\begin{figure}[t]
    \begin{minipage}[b]{1.0\linewidth}
      \centering
      \centerline{\includegraphics[width=6.0cm]{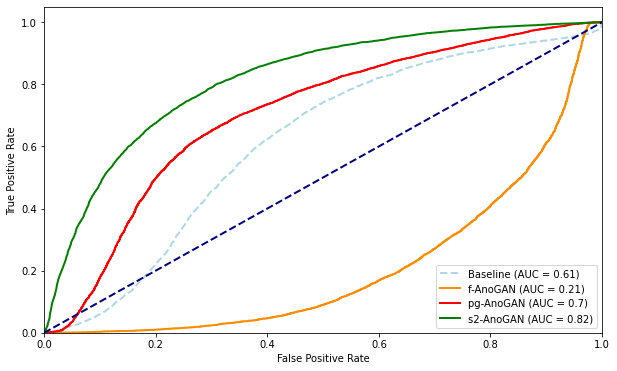}}
      \centerline{(a) ROC curves of anomaly detection}\medskip
    \end{minipage}
   
    \begin{minipage}[b]{0.48\linewidth}
      \def\cpar{\hss\egroup\line\bgroup\hss}
      \centering
      \centerline{\includegraphics[width=4.0cm]{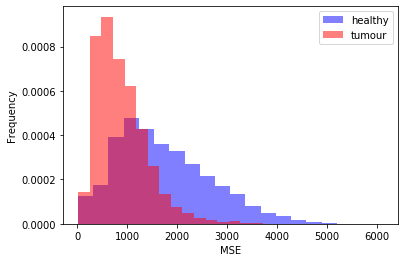}}
      \centerline{(b) Histogram of $\A_{MSE}$}
      \centerline{(f-AnoGAN)}
    \end{minipage}
    \hfill
    \begin{minipage}[b]{0.48\linewidth}
      \def\cpar{\hss\egroup\line\bgroup\hss}
      \centering
      \centerline{\includegraphics[width=4.0cm]{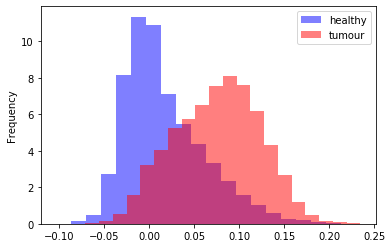}}
      \centerline{(c) Histogram of $\A_{canny}$ }
      \centerline{(s2-AnoGAN)}
    \end{minipage}
    \centering
    \caption{(a) shows ROC curves with the corresponding AUC achieved by the tested frameworks, and the baseline (for which only the number of edges is used as anomaly score). (b) is the histogram of $\A_{MSE}$ scores computed on images and their projections done by f-AnoGAN. (c) is the histogram of anomaly scores computed by our s2-AnoGAN framework.}
    \label{fig_results}
\end{figure}

\subsection{Anomaly detection results} 

\begin{table}[]
    \resizebox{\columnwidth}{!}{%
        \begin{tabular}{l|lll}
                    & f-AnoGAN  & pg-AnoGAN & s2-AnoGAN \\ \hline
        $\A_{f\text{-}AnoGAN}$ & 0.21 & 0.32 & \textbf{0.37} \\
        $\A_{pg\text{-}AnoGAN}$ & \textbf{0.74} & 0.70 & 0.71 \\
        $\A_{Canny}$ & 0.75 & 0.57 & \textbf{0.82}
        \end{tabular}%
    }
        \caption{AUC values of anomaly detection when combined with the anomaly measure of each framework, respectively.}
        \label{tab_results}
\end{table}

 \begin{figure}[t]
    \begin{minipage}[b]{1\linewidth}
      \def\cpar{\hss\egroup\line\bgroup\hss}
      \centering
      \centerline{\includegraphics[width=6.0cm]{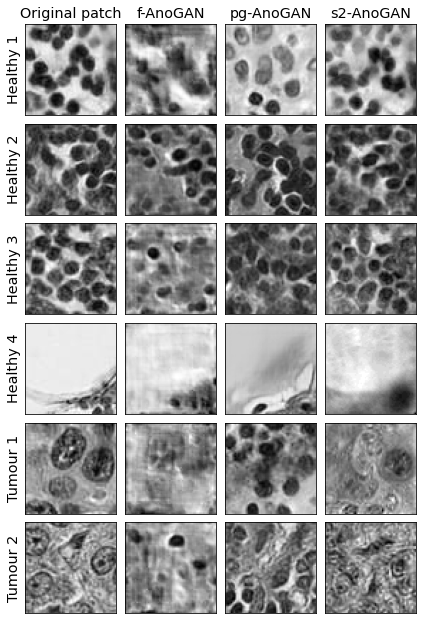}}
    \end{minipage}
    \centering
    \caption{Examples of test data and their projections by f-AnoGAN, pg-AnoGAN and s2-AnoGAN frameworks. }
    \label{fig_reconstructions}
\end{figure}  

We tested the three anomaly frameworks as well as other combinations of GAN+projector and anomaly measure. The performance is measured in terms of AUC (Area Under the Receiver Operating Characteristics Curve). In Table \ref{tab_results}, we see that the performance of $\A_{f\text{-}AnoGAN}$ and $\A_{pg\text{-}AnoGAN}$  on the pathology data is considerably worse than reported in \cite{fAnoGAN} and \cite{am2019unsupervised} for all GAN frameworks. The highest AUC score of 0.82 was achieved by our framework: StyleGAN2 with gradient descent projector and the anomaly metric $\A_{Canny}$. 

The ROC curves for the three frameworks are presented in Sub-figure (a) of Fig.  \ref{fig_results}, further illustrating the top result from s2-AnoGAN. 
As a sanity check, we also include a baseline measure not involving any GAN training: using $N_{x_{Canny}}$ as an anomaly score (number of edge pixels in the input image). We can see that s2-AnoGAN and pg-AnoGAN outperformed the baseline by a large margin but f-AnoGAN actually achieved a lower AUC. 
To gain a better understanding of the results, in the next sections we will  discuss the two main factors affecting the anomaly detection performance: GAN expressivity and the chosen anomaly metric.

 \subsection{Quality of learned underlying distribution}\label{factor_gan}

GANs and their projectors were evaluated based on qualitative as well as quantitative inspection. Here we mainly focus on the healthy data as it was the target of the training. The overall qualitative impression is that s2-AnoGAN was most successful in reproducing realistic images, as illustrated in Fig.~\ref{fig_reconstructions}. From healthy example patches 1-3, we can see how s2-AnoGAN reconstructs typical healthy samples, with several visible cell nuclei, better than the other two frameworks. The amount of cell nuclei and their positions shows a better agreement with the input in these cases.  

This observation is also supported by the peak signal-to-noise ratio (PSNR) between the original healthy image and its reconstruction. Average PSNR score on the healthy data for s2-AnoGAN is 28.7 (standard deviation: 1.7) while for f-AnoGAN and pg-AnoGAN frameworks it is 16.6 (standard deviation: 3.3) and 13.4 (standard deviation: 2.6), respectively. The combination of the highest PSNR mean value and the smallest standard deviation achieved by s2-AnoGAN indicates that the framework will be better suited for a distance based anomaly score.

However, less common types of healthy patches, such as example 4 in Fig.~\ref{fig_reconstructions}, are poorly reconstructed by all three frameworks. For reference, Fig.~\ref{fig_reconstructions} also include two tumour images, showing the expected effect of poor reconstruction.

\subsection{Problem with pixel-wise anomaly measures}\label{factor_metric}

\begin{table}[]
\resizebox{\columnwidth}{!}{%
\begin{tabular}{l|lll}
            & f-AnoGAN  &  pg-AnoGAN & s2-AnoGAN \\ \hline
$ LPIPS $ & 0.36 & 0.36 & \textbf{0.64}  \\
$\A_{D}$ & 0.24 & 0.35 &  \textbf{0.36} \\
$\A_{MSE}$ & 0.21 & 0.32 &  \textbf{0.37} \\
$\A_{res}$ & 0.19 & \textbf{0.33} &  0.27 \\
$\A_0(z)$ & \textbf{0.75} & 0.70 & 0.71 \\

\end{tabular}%
}
\caption{AUC values achieved by different anomaly measures. The latter four are components of the measures used by f-AnoGAN and pg-AnoGAN.}
\label{tab_results_other}
\end{table}

An interesting result is that the pixel-wise difference measures work poorly for anomaly detection in pathology data. From Fig.~\ref{fig_results}(b), we can see that the anomaly score using $\A_{MSE}$ actually correlated with healthy tissue, the opposite of what was expected. In theory, this indicates that the projections of tumour patches are better than the healthy. The same effect is seen for the related $\A_{res}$ score (see Table \ref{tab_results_other}). 
The explanation is two-fold. The MSE values are based on the locational comparison of the pixels, hence, it can be high even when the result is realistic. For example, when the exact positions of cell nuclei turn up to be slightly moved in the reconstructed images (see Fig.~  \ref{fig_reconstructions}). Conversely, the reconstructions of tumour tissue can have the characteristics of a blurred image, giving relatively low MSE even though the realism is poor.

To overcome the problems with the direct comparisons between input and reconstructed pathology images, we introduced the $\A_{Canny}$ anomaly score. We also investigated other distance functions not affected by the exact locations of pixels, including LPIPS~\cite{LPIPS}. LPIPS is a recent learning-based measure that has shown good performance in image comparison problems. LPIPS achieved an AUC of 0.64 (see Table \ref{tab_results_other}) when combined with StyleGAN2 and its projector versus AUC of 0.82 when using $\A_{Canny}$ (see Table \ref{tab_results}). This is likely due to LPIPS being based on features learned by a neural network trained on natural images, which seems to be a poor representation of the important features in pathology data.

\section{Conclusions and Future Work} \label{sec:conclusion}

In this work we explored unsupervised anomaly detection with GANs and how it works on histopathology data. Due to the more complex structures in the images than used in previous work, the current methods did not perform well and were outperformed by our introduced s2-AnoGAN framework with anomaly score based on canny edges. One  limitation of the StyleGAN2 projector is that it is based on gradient descent. Thus, it is much slower than the projection methods used by Schlegl et al. \cite{fAnoGAN} and Berg et al. \cite{am2019unsupervised}. For future work, we will explore different encoder-based projectors and tailor them to work well with the StyleGAN2 architecture. Furthermore, we will explore if the results could be further improved by deriving other metrics for comparing original pathology images with their reconstructions.

\section{Acknowledgments}
\label{sec:acknowledgments}
This work was supported by the Swedish e-Science Research Center and VINNOVA, grant 2017-02447. Claes Lundstr\"om is an employee and shareholder of Sectra AB.

\bibliographystyle{IEEEbib}
\bibliography{strings,refs}

\section{Compliance with Ethical Standards}
This research study was conducted retrospectively using human subject data made available in open access by Litjens et al \cite{camelyon17}. Ethical approval was not required as confirmed by the license attached with the open access data.

\end{document}